# Programmable thermocapillary shaping of thin liquid films

Ran Eshel[1,†], Valeri Frumkin[1,†,§], Matan Nice[1], Omer Luria[1], Boris Ferdman[2,3], Nadav Opatovski[2], Khaled Gommed[1], Maxim Shusteff[4], Yoav Shechtman[2,3], and Moran Bercovici[1,2,3,*]

[1] Faculty of Mechanical Engineering, Technion - Israel Institute of Technology, Haifa, 3200003 Israel

[2] Department of Biomedical Engineering, Technion - Israel Institute of Technology, Haifa, 3200003 Israel

[3] Russell Berrie Nanotechnology Institute, Technion - Israel Institute of Technology, Haifa, 3200003 Israel

[4] Lawrence Livermore National Laboratory, 7000 East Ave, Livermore, CA 94550, USA

§ Current affiliation: Department of Mathematics, Massachusetts Institute of Technology, Cambridge, MA, 02139, United States

†The author equally contributed

*Corresponding author: M.B. (*mberco@technion.ac.il*)

**Abstract**

We present a method that leverages projected light patterns as a mechanism for freeform deformations of a thin liquid film via the thermocapillary effect. We developed a closed-form solution for the inverse problem of the thin-film evolution equation, allowing to obtain the projection pattern required in order to achieve a desired topography. We experimentally implement the method using a computer controlled light projector, which illuminates any desired pattern onto the bottom of a fluidic chamber patterned with heat-absorbing metal pads. The resulting heat map induces surface tension gradients in the liquid-air interface, giving rise to thermocapillary flow that deforms the liquid surface. If a polymer is used for the liquid film, it can then be photocured to yield a solid device. Based on the inverse problem solutions and using this system, we demonstrate the fabrication of several diffractive optical elements (DOEs), including phase masks for extended depth of field imaging, and for 3D localization microscopy. The entire process, from projection to solidification, is completed in less than five minutes, and yields a sub-nanometric surface quality without any post-processing.

**Impact Statement**

The ability to arbitrarily control the topography of a thin liquid film can be beneficial for a range of applications, from optics to biology. However, no methods exist today for achieving programmable surface deformations. This work shows that photoactuation, achieved using a low intensity projection, can effectively drive thin film deformations via the thermocapillary effect. Furthermore, the inverse problem approach to the thin film equation and the analytical solutions that we present here bridge the gap between science and engineering by providing the illumination pattern required in order to produce a desired topography. Lastly, the method enables rapid prototyping of diffractive optical elements - answering an unmet need in the optical design industry.

# 1. Introduction

Thin liquid films are ubiquitous throughout the natural world and play important roles in a wide range of technological processes. For example, they can serve as a transport mechanism in microfluidic devices (Stone, Stroock, & Ajdari, 2004), as liquid substrates for colloidal self-assembly (Nagayama, 1996), or in cooling systems (Kabov, Gatapova, & Zaitsev, 2008); they can also be used for fabricating optical micro-lens arrays (Chronis, Liu, Jeong, & Lee, 2003; Moran et al., 2006), and play an important role in many biological systems such as in the lining of the lungs and eyes (Grotberg, 1994; Wong, Fatt, & Radke, 1996). It is therefore of significant importance to develop robust and precise methods for controlling the behavior of thin liquid films and for manipulating the topography of their free surface.

Several mechanisms for thin film manipulation have been demonstrated over the past years. One such approach was demonstrated by Brown et al who used dielectrophoretic forces produced by a set of interdigitated electrodes to create periodic deformation on the interface of a thin layer of polymer, thus solidify them into diffraction gratings (Brown, Wells, Newton, & McHale, 2009; Wells, Sampara, Kriezis, Fyson, & Brown, 2011). Recently, Gabay et al. (Gabay et al., 2021) extended this concept to allow greater control over the deformation topography using more complex electrode patterns. However, this approach requires pre-patterning of electrodes using complicated and expensive lithography processes.

An alternative actuation mechanism for manipulation of liquid interfaces is the use of the Marangoni effect, where variations in surface tension yield mass transfer across the liquid interface, resulting in its deformation. Several works demonstrated photochemical manipulation of surface tension to pattern thin polymer films (Katzenstein et al., 2012; Kim, Janes, Zhou, Dulaney, & Ellison, 2015). However, similarly to the dielectrophoretic approach, photochemical manipulation of surface tension requires preparation of specialized photomasks, thus preventing arbitrary manipulation of the free surface. A potentially simpler way to induce variations in surface tension is by means of the thermocapillary effect. Numerous studies have considered the use of this approach for patterning thin polymer films (Singer, 2017). In particular, Nejati et al. (Nejati, Dietzel, & Hardt, 2016) leveraged the Benard-Marangoni instability to drive short-wavelength deformation in a polymer film, yet their approach was limited to periodic structures. McLeod and Troian demonstrated deformations of nanoscale films using temperature-controlled structural elements brought in close proximity to the film, but this required the mechanical fabrication of conducting metal structures for each desired deformation (McLeod, Liu, & Troian, 2011). Programmable and dynamic freeform manipulation of liquid interfaces remains an important open problem in the thin film community.

In this work, we present for the first time the use of projected light patterns as a mechanism for driving freeform deformations of a thin liquid film via the thermocapillary effect. As illustrated in Figure 1, we use a standard programmable digital micromirror device (DMD) to project a desired illumination intensity map onto a substrate containing an array of light-absorbing metal pads. This provides unprecedented control over the spatial temperature distribution in the substrate, which is translated to surface tension gradients and subsequent surface deformations in the overlaying liquid film. We derive a theoretical model for an inverse problem that provides the required temperature map for achieving a desired topography. When using a curable polymer, the resulting topography can be solidified to create specially designed molecularly smooth surfaces. The entire shaping and curing process is completed in under 5 minutes, and the same light-absorbing substrate can be used again to create a different topography using a new projected pattern.

We chose to demonstrate the applicability of our method for rapid prototyping of freeform diffractive optical elements (DOEs). DOEs are used to manipulate light by introducing non-uniform phase accumulation through a carefully designed geometry and/or by changing the refractive index of the optical element (Born & Wolf, 2013), and are key components a in a wide range of optical applications (Turunen & Wyrowski, 1998). DOEs are typically created using clean room fabrication techniques, such as direct writing, focused ion beam milling, grayscale lithography and interferometric exposure ("Diffractive Optics: Design, Fabrication, and Test | (2003) | O'Shea | Publications | Spie," n.d.), or by precise multi-axis mechanical processes such as diamond machining or magnetorheological finishing (MRF) (Kordonski, 2011; Rolland et al., 2021). However, all of these processes require significant and expensive infrastructure and highly-skilled operators, which lowers accessibility and drives high component costs. Furthermore, existing techniques do not allow for rapid prototyping which is much desired for driving research and design innovation.

Here, using the inverse-problem approach, we first demonstrate the fabrication of a one-dimensional diffraction grating component and characterize its performance, showing good agreement with theory. We then demonstrate the fabrication of an axisymmetric phase mask for extended depth of focus (EDOF) and show its implementation in imaging over a greater focal depth without reducing the numerical aperture. Finally, we demonstrate the fabrication of a non-axisymmetric saddle-point phase mask (Shechtman, Sahl, Backer, & Moerner, 2014), and experimentally demonstrate its use in encoding of depth information in the point spread function (PSF) of nanoparticles, allowing microscopic three-dimensional particle tracking (Ferdman et al., 2020). We provide complete details of the experimental system, and characterize its resolution and dynamic range as a function of the illumination intensity and of the absorbing array geometry. Our method thus provides a potential solution for low-cost customized fabrication of DOEs for industrial applications, and allows for rapid prototyping of new DOE-based ideas in research.

## 2. Results

### 2.1. Principles of the method

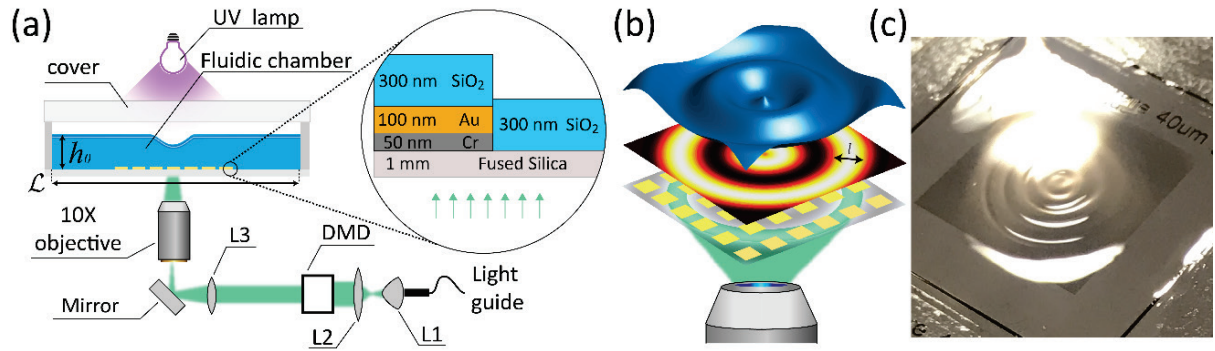

*Figure 1. Operational principle of the thermocapillary fluidic shaping method.* (a) Schematic illustration of the experimental setup. The setup is based on a shallow fluidic chamber filled with a thin layer of a curable polymer. The bottom surface of the chamber is a glass substrate patterned with an array of metal pads designed to absorb light in the visible spectrum. A desired illumination pattern is projected onto the surface using a DMD-based system. The inset describes the internal structure of the metallic pads. (b) A desired illumination pattern is projected onto the metal pads, which absorb the light and create a corresponding temperature field. Heat is transferred from the pads, through the thin liquid layer and to the liquid-air interface, leading to surface tension gradients that drive the thermocapillary effect, resulting in spatial deformations of the free surface. (c) Image of the solidified polymer after exposure to UV illumination.

Figure 1 presents the principle of operation of the thermocapillary fluidic shaping method and the experimental setup used for its implementation. When localized heating is imposed on a gas-liquid interface, the liquid will flow outward along the free-surface from warmer regions with lower surface tension to colder regions with higher surface tension. This results in local deformation of the liquid film (Tan, Bankoff, & Davis, 1990), and is known as the thermocapillary effect. The magnitude of the surface deformation is proportional to the local temperature gradients. We found that a convenient way to control these temperature gradients is to use light projection, whose spatial intensity can be easily programmed, to heat an array of absorbing metal pads patterned on the bottom of the fluidic chamber. The system uses a standard < 3W microscope LED source and a standard DMD. The resulting increase in temperature is on the order of few degrees or less, and is sufficient to drive significant deformations.

The spatial resolution of the thermal patterning depends on the material properties of the substrate, the resolution of the projected pattern, and the geometry of the light absorbing metallic pad array. The composition of the metal pads was designed to maximize light absorbance across the UV-Vis spectrum of the light source, as shown in Figure 1c. Section 2 in the Supplementary Information provides characterization of the deformation magnitude and resolution in our system as a function of the illumination intensity and pad array density. As expected, the magnitude of the resulting deformation is proportional to both the illumination intensity and the area fraction occupied by the pads. The precise deformation magnitude and resolution depend on the specific heating pattern, with some tradeoff between the two (see theory section). However, typical values are on the order of 3 μm maximum deformation and 5 μm/mm spatial resolution.

## 2.2. The inverse problem — theory and implementation

Consider an open microfluidic chamber of length $\mathcal{L}$, containing a thin liquid film of thickness $h_0$, as depicted in Figure 1(a). The chamber is heated from below by a light pattern that is absorbed by the metal pads array, resulting in temperature variations over a characteristic length $l$, $h_0 \ll l \ll \mathcal{L}$. For small variations in temperature, we can assume that the surface tension, $\sigma$, changes linearly with temperature, $\theta$, according to

$$\sigma = \sigma_r - \beta(\theta - \theta_r), \tag{1}$$

where $\beta \equiv \frac{\partial \sigma}{\partial \theta}$ is defined as the thermocapillary coefficient and is assumed to be constant, $\theta_r$ is a reference temperature, and $\sigma_r$ is the surface tension at that temperature. We assume that variations in temperature are sufficiently small such that changes in density and viscosity are negligible.

In Supplementary Information section 3 we provide a complete derivation of the governing equations. Briefly, we consider the continuity equation, the steady-state incompressible Navier-Stokes equations, and the steady-state advection-conduction heat equation. At the bottom of the film ($z = 0$) we require no-slip and no-penetration conditions on the flow, and set a fixed temperature distribution, $\theta_b$. At the liquid air interface ($z = h(x, y)$) we impose the kinematic condition relating flow velocities to the shape of the deformed interface, a stress balance relating the stress tensor in the liquid to surface tension, and heat loss through Newton's cooling law. Using these equations and boundary conditions, we derive the dimensionless steady-state equation of the deformed liquid film driven by a non-uniform temperature distribution,

$$\vec{\nabla}_2 \cdot \left[ \frac{S}{3} H^3 \vec{\nabla}_2 \nabla_2^2 H - \frac{G}{3} H^3 \vec{\nabla}_2 H - \frac{1}{2} H^2 \vec{\nabla}_2 \left( \frac{\vartheta_b + \Theta}{1 + BH} \right) \right] = 0. \tag{2}$$

Here $[X, Y, Z, H] = \left[ \frac{x}{l}, \frac{y}{l}, \frac{z}{h_0}, \frac{h}{h_0} \right]$ are the dimensionless coordinates and film thickness scaled by the characteristic lateral and vertical dimensions, and $\vec{\nabla}_2 = \left( \frac{\partial}{\partial X}, \frac{\partial}{\partial Y} \right)$ and $\nabla_2^2 = \left( \frac{\partial^2}{\partial X^2} + \frac{\partial^2}{\partial Y^2} \right)$ are the two-dimensional gradient and Laplacian operators. We scale the velocity as $[U, V, W] = \frac{1}{U_0} [u, v, \frac{1}{\varepsilon} w]$, where $U_0 = \varepsilon \Delta \sigma / \mu$ is the characteristic velocity as derived from the tangential stress balance, with $\Delta \sigma = \beta \Delta \theta$ being the characteristic variation in surface tension, and $\Delta \theta = \theta_{b_{max}} - \theta_{b_{min}}$ the difference between the maximum and minimum temperature at that surface. $\mu$ is the dynamic viscosity of the liquid, and $\varepsilon = \frac{h_0}{l} \ll 1$ is the ratio of film thickness to characteristic lateral scale. We scale the temperature as $\vartheta = \frac{\theta - \overline{\theta}_b}{\Delta \theta}$ where $\overline{\theta}_b$ is the average temperature of the bottom surface. $S = \varepsilon^3 \frac{\sigma_r}{\mu U_0} = \varepsilon^2 \frac{\sigma_r}{\Delta \sigma}$ is the surface tension number, $B = \frac{\hbar h_0}{k}$ is the Biot number, $\hbar$ is the heat transfer coefficient at the liquid-air interface, $k$ is the thermal conductivity, $\Theta = \frac{\overline{\theta_b} - \theta_\infty}{\Delta \theta}$ is a ratio of temperature differences where $\theta_\infty$ is the ambient temperature, and $G = \varepsilon \frac{\rho h_0^2}{\mu U_0} g = \frac{\rho h_0^2}{\Delta \sigma} g$ is the dimensionless gravity.

Subjected to appropriate boundary conditions on the thickness of the liquid film (e.g. far from the heating region, the film thickness can be assumed to be known and uniform), equation 2 can be solved numerically to obtain a solution to the direct problem - the spatial deformation resulting from any prescribed temperature distribution $\vartheta_b(x, y)$. However, for engineering purposes, the inverse problem —seeking the temperature map that would yield a desired deformation — is of greater value. As shown in Supplementary Information section 3, for sufficiently small Gaussian curvatures, and assuming that far from the heating region the liquid film is flat, the temperature distribution can be obtained explicitly as a function of the desired deformation, $H(x, y)$,

$$\vartheta_b = \frac{(1 + BH)}{3}\left[2SH\nabla^2 H - S(\vec{\nabla}H)^2 - G(H^2 - 1)\right] - \Theta. \qquad (3)$$

Figure 2 presents the use of the inverse problem for fabricating a desired topography. Figure 2a presents the prescribed topography composed of a set of concentric rings, with a maximum deformation relative to the baseline film thickness of $d_{desired} = 1.5\ \mu m$. This topography is used as input into equation (3), with the constant parameters as defined in SI section 3, resulting in a temperature distribution $\vartheta_b(x, y)$. Since we do not have a direct relation between desired temperature and the required illumination intensity, we assume a linear dependence and perform a calibration measurement in which the illumination intensity map is set to $I(x, y) = I_0\ \vartheta_b(x, y)/\max\{\vartheta_b(x, y)\}$, where $I_0$ is the maximum intensity of the light source. The resulting intensity map is presented in Figure 2b, and an image of the actual projected intensity map is presented in Figure 2c. We measure the resulting maximum deformation, $d_{max}$ (which for this pattern was 2.2 um), and repeat the projection using an intensity map that is scaled by $d_{desired}/d_{max}$ (i.e. 1.5/2.2). Figure 2d presents the resulting topography of the (solidified) deformed liquid film.

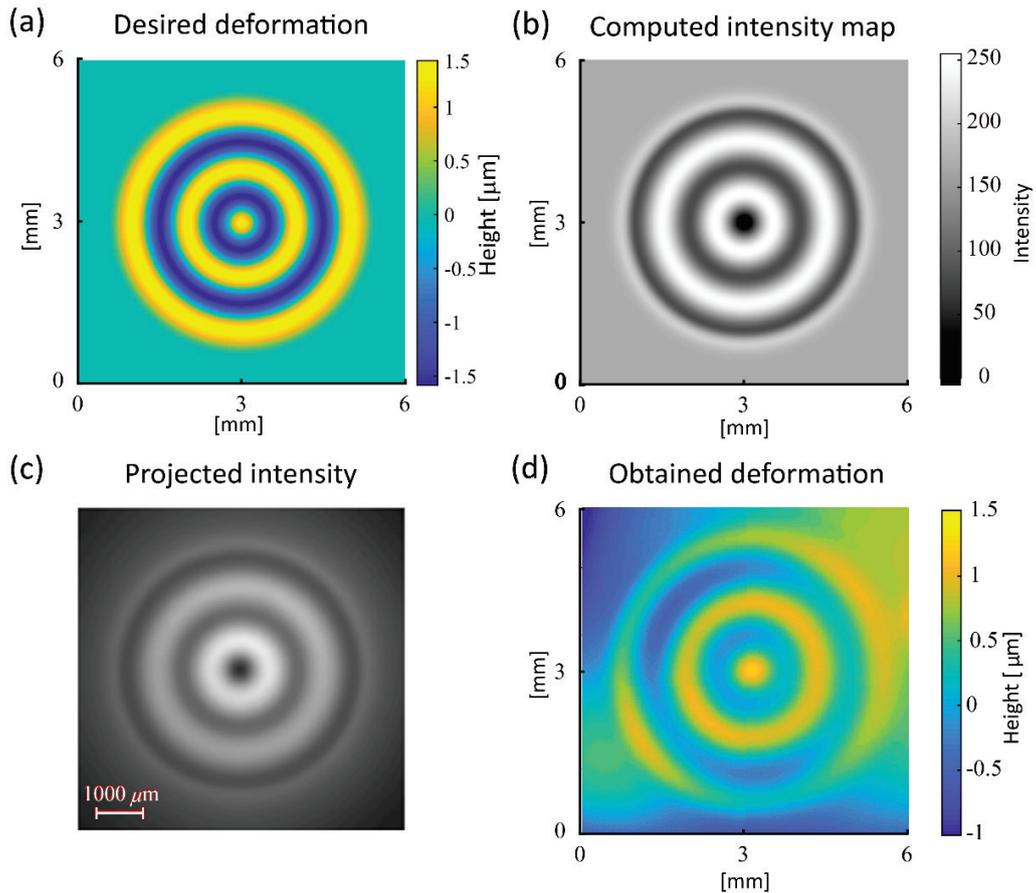

*Figure 2. Demonstration of a complete workflow for an inverse problem solution. (a) The desired topography defined in terms of the variation from a reference height. (b) The intensity map required in order to produce the desired deformation, as obtained from solving the inverse problem. (c) Image of the illumination pattern as obtained on the bottom surface of the fluidic device. (d) The resulting topography after curing the polymer as measured by the Digital holographic microscope (DHM) (Cuche, Marquet, & Depeursinge, 1999).*

**2.3. Applications**

Figures 3-5 present several practical diffractive elements produced using the thermocapillary fluidic shaping approach. Diffraction gratings are optical elements with a periodic structure, that diffract the incident light into several beams traveling in different directions, depending on the wavelength and the spatial periodic structure defining the angular spread. They are commonly used in monochromators for spectroscopic instruments (Loewen & Popov, 2018).

Figure 3 presents a linear diffraction grating with a peak-to-peak periodicity of 400 μm produced by projection of a sinusoidal illumination pattern. Figure 3a presents the projected pattern side-by-side with the resulting deformation, and Figure 3b presents the magnitude of the deformation along the centerline. We test the resulting element by measuring the diffractive angle for several wavelengths produced by monochromatic LEDs. Since the LEDs have a spectral bandwidth of approximately 50 nm, we use a weighted average of their spectrum (as provided by the manufacturer) to define a representative wavelength for the analysis. As shown in Figure 3c, the results are in excellent agreement with the expected angle set by the element's periodicity. Figure 3d also presents a visual demonstration of white light diffracted through the grating and directly imaged by a camera sensor, clearly showing the separation of bands.

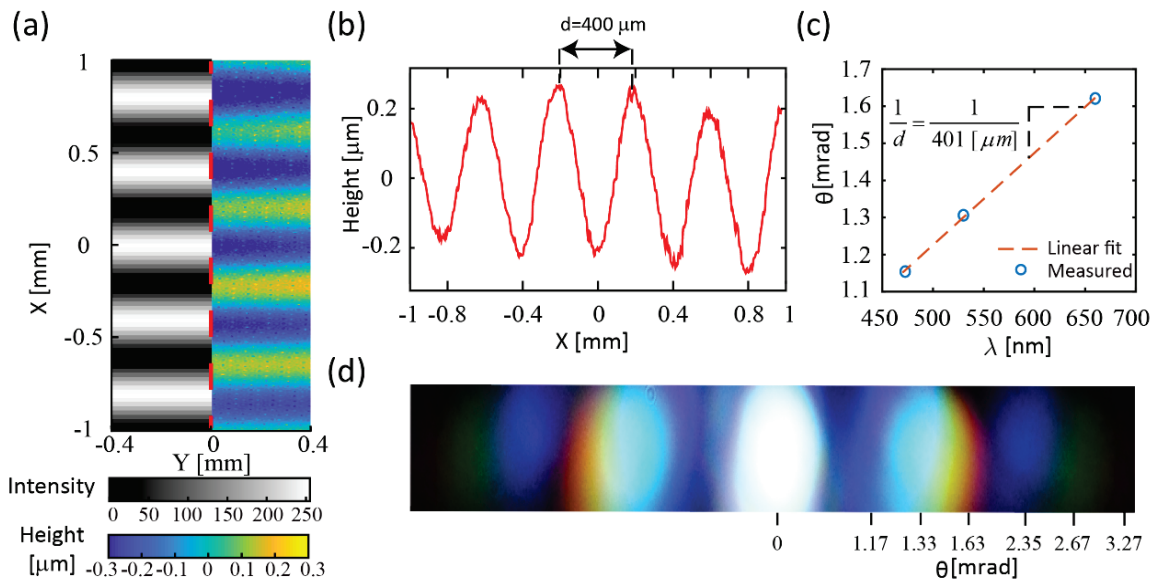

*Figure 3. Fabrication and testing of a linear diffraction grating. (a) Side-by-side image showing the projected light pattern (left) and the measured resulting topography (right). Regions with a higher projected intensity, and therefore a higher temperature, correspond to valleys in the topography. (b) The deformation magnitude along the center of the element, showing the spatial frequency (400 μm) and the resulting amplitude (0.4 μm) of the grating. (c) The measured diffractive angle for three wavelengths passing through the grating. The reciprocal of the slope of a least-square line matches well the designed grating spatial frequency. (d) Image of white light diffraction. The numbers along the scale indicate the center of each band along a horizonal centerline.*

The depth of field of any optical system can be extended by reducing its numerical aperture, yet at the cost of reducing the amount of light collected. A number of studies have shown the ability to overcome this limitation and extend the depth of field without loss of photons, by positioning an appropriate phase mask in the systems' aperture (Ben-Eliezer, Marom, Konforti, & Zalevsky, 2005; Dowski & Cathey, 1995). Figure 4 presents the fabrication and testing of an extended depth of field (EDOF) phase mask designed by formulating a phase retrieval task as described by Nehme *et al.* (Nehme et al., 2021). The desired mask mold is translated into a light pattern (Figure 4b) using the inverse problem solution, and the thermocapillary fluidic shaping method is used to create a solid mold. We then cast PDMS onto the mold to create the phase mask itself, shown in Figure 4c. We position the mask in the Fourier plane of a *4f* system and use it to image three targets positioned at different distances from the camera. As illustrated in Figure 4d, to ensure that the presence of the mask itself does not reduce the numerical aperture, we use a smaller fixed aperture in close proximity to the mask, acting as the aperture stop of the system. Figures 4e-f clearly show the additional depth of field gained by using the mask, yet with some loss of contrast.

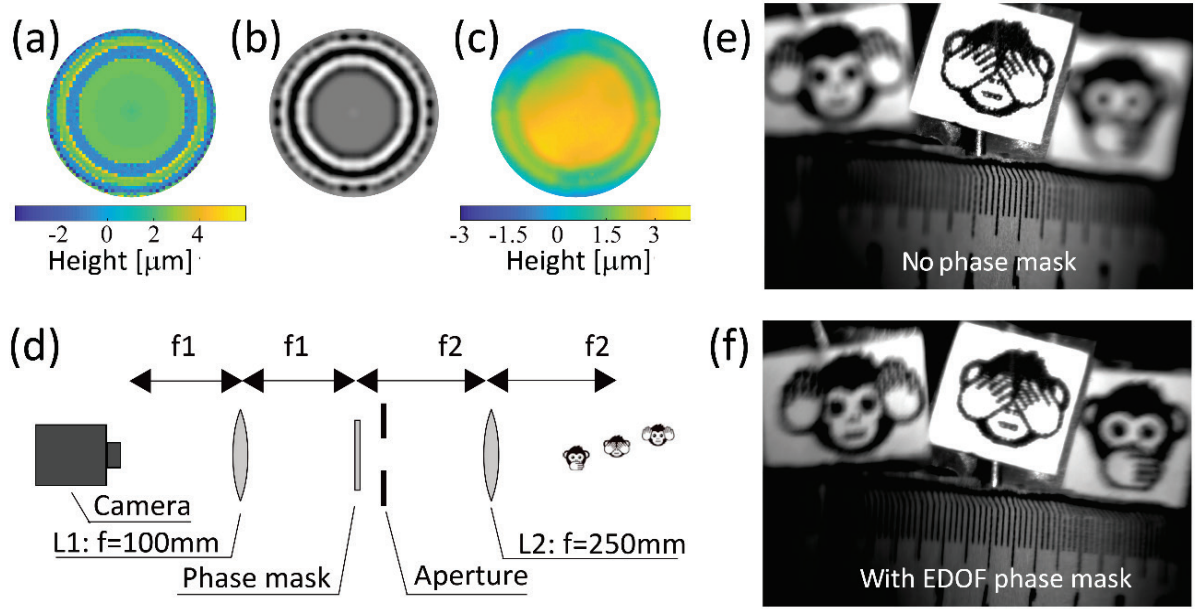

*Figure 4. Fabrication and demonstration of an extended depth of field phase mask.* *(a) The design of the extended depth of field (EDOF) mask based on Nehme et al. (Nehme et al., 2021). (b) The projection pattern for a negative mold, as solved by the inverse problem. (c) The obtained phase mask as measured by the DHM. (d) Schematic illustration of the optical setup. We position the phase mask in the Fourier plane of a 4f system, and image three targets positioned 2.5 cm from one other. We used an additional small fixed aperture to ensure that the phase mask itself does not reduce the aperture stop. (e) The resulting image without the phase mask, showing only the center target in focus. (j) The resulting image with the EDOF phase mask showing all three targets in focus.*

Three-dimensional localization microscopy uses phase masks positioned in the Fourier plane of a microscope to modify the PSF in order to encode three-dimensional information (Pavani et al., 2009). Figure 5 presents the use of the inverse problem approach for fabrication of such a phase mask. Figure 5a presents the topography of the desired 'tetrapod' mask (Shechtman et al., 2014), designed to provide depth information over a range of +/- 2 μm. Here, too, we use the desired mold topography as input to equation (3) and obtain the intensity map presented in Figure 5b. The parameters used in the solution are identical to those that were used in Figure 4 and that are detailed in SI section 3. Figure 5c presents the topography of the resulting phase mask created by casing PDMS onto the resulting mold, as measured by the DHM. The first column in Figure 5d presents the measured PSF of a single emitter (a 200 nm fluorescent bead) at different locations along the optical axis, in the absence of any phase mask. The second column presents the measured PSF of the same emitter, when using a phase mask produced using a 3-step lithography process. The last column in Figure 5d presents the measured PSF as obtained using our phase mask fabricated by the thermocapillary fluidic shaping method. Figure 5e demonstrates the use of this mask for 3D tracking of three 200 nm beads undergoing Brownian motion (see Supplementary Video 1) in a liquid environment, and of one stationary bead (attached to the bottom surface).

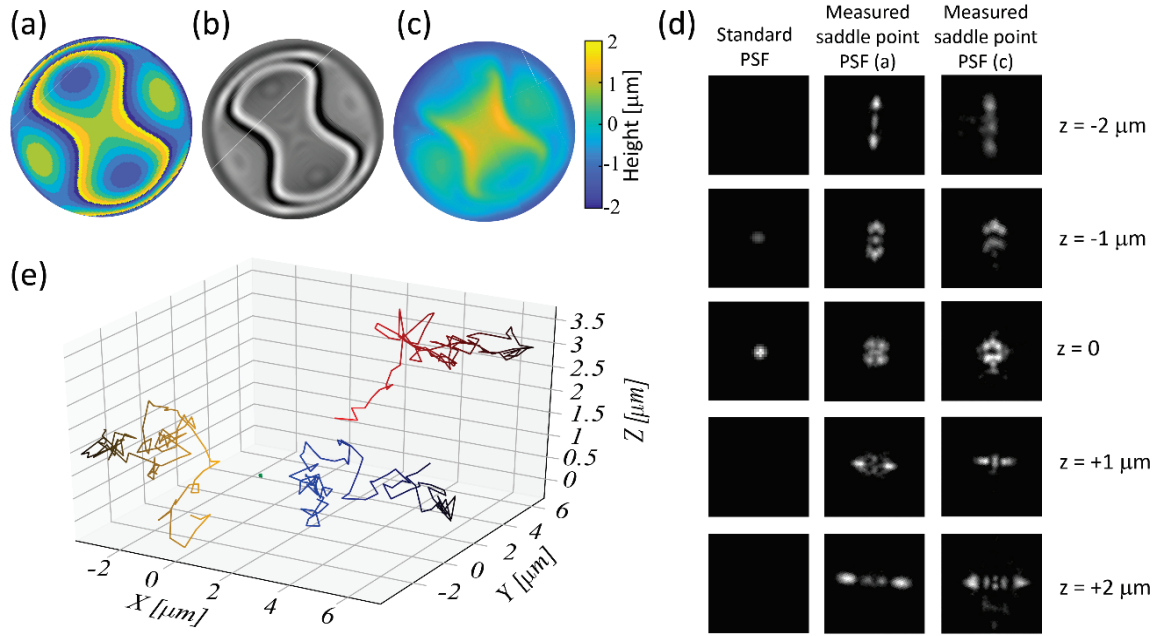

*Figure 5. Fabrication and demonstration of a saddle point phase mask for three-dimensional localization microscopy.* **(a)** *Topography map of a saddle point phase mask microfabricated using standard lithography and ion etching processes.* **(b)** *The illumination pattern required for deforming the liquid film into a negative mold for the phase mask, as obtained from the inverse problem solution.* **(c)** *DHM measurement of the resulting solidified PDMS mask, cast on the mold fabricated by thermocapillary fluidic shaping.* **(d)** *The PSF of a 200 nm diameter fluorescent microsphere at various z positions, as obtained from (i) a measured standard PSF with no phase mask (ii) a measured PSF using a lithography phase mask, (iii) a measured PSF using the mask obtained through thermocapillary fluidic shaping.* **(e)** *Tracking of three nanobeads undergoing Brownian motion and one stationary bead (in green) in a liquid, using the fabricated mask.*

## 4. Conclusion

In this work, we introduced a novel method for thermocapillary shaping of thin polymer films, and applied it as means for the rapid fabrication of DOEs. We provided a theoretical model for the inverse problem, making it possible to design specific topographies corresponding to desired optical functionalities. This approach has several significant advantages over existing fabrication technologies. First and foremost, the method is programmable in the sense that the desired deformation is created in response to the projected pattern. Thus, a single setup can be used to implement a wide range of different designs. Once the liquid element is exposed to UV light, it undergoes rapid solidification (in under 5 minutes) and results in a molecularly smooth solid element. The short fabrication time scale makes the method especially attractive for rapid prototyping of DOE's, allowing maximal flexibility for testing various optical designs. To illustrate the utility of the method, we fabricated several DOE's and demonstrated their functionality.

Using our current setup, we were able to achieve deformations on the order of several microns over a lateral span of millimeters. The magnitude of deformations is sufficient for most applications in the visible spectrum, as it allows for phase accumulation of at least $2\pi$. In the relatively thick films considered here, spatial resolution is limited by the natural heat convection from the liquid film. Higher spatial resolution could potentially be achieved by working with thinner films, as was previously demonstrated in other works on Marangoni-based patterning (Katzenstein et al., 2012; Kim et al., 2015). Additional techniques such as effective heat removal and precisely controlled ambient temperature, will also aid in increasing the spatial resolution of the method.

Our inverse problem solution provides the temperature distribution required for a desired deformation. In our implementation, we assume a linear relation between the desired temperature and the projected light intensity. However, this assumption must be verified, and may be the source of deviation between the desired and obtained topographies. More generally – an empirical model relating the temperature to projection intensity should ideally be created. To do so, the temperature on the liquid interface should be measured, requiring high resolution and high precision thermal imaging. Alternatively, the topography of the liquid film could be measured in real-time (e.g. by interferometry) allowing to close a loop and modify the projection pattern dynamically to match the desired shape before the element is solidified. Another attractive feature of this approach is the simplicity and affordability of the required setup, consisting of a simple projector and a reusable substrate containing an array of light-absorbing metal pads. As such, this technology can be easily scaled up and implemented in any laboratory, allowing for low-cost customized fabrication of DOEs for both research and industrial applications.

Finally, while in this work we focused on fabricating solid topographies, the shape of the liquid interface can be dynamically controlled by varying the projected light in real time. This capability suggest that the method presented here could potentially be applied to fields such as adaptive optics, colloidal assembly, and manipulation of cells and particles in the biomedical sciences.

## Appendix A: Methods and materials

### Thermocapillary shaping experimental setup

The experimental setup consists of a glass substrate patterned with metallic pads fabricated by liftoff photolithography processes. The pads consist of two metallic layers - a 150 nm chromium layer, followed by a 50 nm gold layer. The metal pads were then coated with a 400 nm silicon dioxide layer to protect them. We used PVC tape to form the boundaries of a 2 cm x 2 cm x 60 $\mu$m fluidic chamber on top of the patterned substrate. Before filling the fluidic chamber, we treated the device with low-pressure air plasma (Zepto W6, Diener Electronics, Germany) for 1 min, allowing for improved wetting of the substrate. Next, we injected ~15 ul of a UV curable polymer (CPS 1050 UV, Colorado Photopolymer Solutions, USA) to fill the chamber. We covered the chamber with a UV-transparent polystyrene lid to protect the polymer from contamination. The illumination system (depicted in figure

1) consists of a liquid guide connected to a computer-controlled LED light source (Sola SE, Lumencore, USA) and a beam expander composed of two lenses: L1, an aspherical condenser (ACL25416U-A, Thorlabs, USA) and L2, a biconvex lens (LB1723-A, Thorlabs, USA) to magnify the beam. We collimate the light into a digital mirror device (1024 x 768 pixels,V7001, Texas Instruments, USA), allowing the projection of any desired shape. The projected light passed through L3, a Kohler lens with f=300 mm (AC254-300-A, Thorlabs, USA), which focuses the light onto the back focal plane of the microscope objective (PlanApo, 10x, NA=0.45, Nikon, Japan). To create a desired deformation, we illuminate the bottom of the device with a corresponding light pattern and allow 1 minute for equilibration. We then position the UV lamp (24 W, λ=370 nm) at a distance of ~20 mm above the substrate and illuminate for 4 min to cure the polymer into a solid element. We measure the topography of the resulting components using a digital holographic microscope (DHMR1003, LynceeTec, Switzerland).

**Fabrication of PDMS based DOEs**

We fabricated our DOEs by casting Polydimethylsiloxane (PDMS) at a ratio of 1:10 (cross-linker to resin) on top of the solidified UV polymer, after treating the polymer surface with Chlorotrimethylsilane (Sigma Aldrich, Germany). After casting the PDMS, we placed the mold in a vacuum chamber for 3 hours for degassing, followed by 3 hours at $60^0$ **C** in the oven for curing.

**Engineered point spread function measurement and bead tracking**

We placed the PDMS phase mask in the Fourier plane of a *4f* system, located on the imaging exit of an inverted microscope (Ti2, Nikon, Japan) equipped with a 100x objective (CFI SR HP Plan Apochromat Lambda S 100XC Sil, Nikon, Japan). To create the reference z-stack, we deposited 200 nm diameter fluorescent emitters (TetraSpeck, Thermo Fisher, USA) on a glass slide and excited them with a combination of a 488 nm and 532 nm lasers. We then imaged the point spread functions (PSFs) of the optical system containing the tetrapod phase mask, as a function of the z coordinate of the beads over an axial range of 4.5 μm with steps of 50 nm. The z-stack was used to estimate the optical system's pupil function using phase retrieval [24].

We then placed the same type of beads in a water-glycerol mixture, allowed to move freely under Brownian motion, and imaged them using the tetrapod phase mask. We then used the z-stack to retrieve their three dimensional coordinates, using a maximum likelihood estimation, per frame.

**Diffraction grating measurement**

To measure the diffraction through the fabricated grating, we transmitted collimated monochromatic light through the grating and measured the diffraction pattern on a camera sensor (D90, Nikon, Japan) located at a distance of 1 m. We repeated this using three different LEDs sources (M660F1, M530F1, M470F1, Thorlabs, USA) with spectral peaks at 660 nm, 530 nm and 470 nm. The three LEDs were fiber coupled, allowing the use of a single collimation setup, and switching of the light sources by connecting the fiber input to the different sources. The collimation setup was based on an aspheric condenser (ACL25416U-A, Thorlabs, USA) focused onto a 50 μm pinhole, and 2x objective (Plan UW, NA 0.06, Nikon, Japan) for collection of the light.

**EDOF measurement**

The imaging setup is presented in figure 5d. The *4f* imaging system consists of two lenses, L1 (AC254-100-A, Thorlabs, USA) and L2 (AC254-250-A, Thorlabs, USA), and a monochromatic CCD sensor. The phase mask was placed at the Fourier plane of the system, and an aperture was located adjacent to it. In the reference experiment, the phase mask was removed, but the aperture was maintained, to ensure the same aperture stop for both cases. The imaged target consisted of three of three ~5 mm printed figures connected to a ruler and located at distances of 247.5 mm, 250 mm, and 252.5 mm from the L2 lens.

Acknowledgments. We thank Baruch Rofman for performing the AFM measurements for surface roughness characterization.

Funding statement. This project has received funding from the European Research Council (ERC) under the European Union's Horizon 2020 Research and Innovation Programme, Grant Agreement No. 678734 (MetamorphChip).

Competing interests. The authors declare that they have no competing interests.

Data availability statement. All relevant data is contained within the manuscript.

Ethical standards. The research meets all ethical guidelines, including adherence to the legal requirements of the study country.

Author contributions. V.F and M.B. conceived the method. R.E. developed the experimental setup, performed the thermocapillary experiments, collected and analyzed the data. V.F. and M.N. developed the theory. R.E., V.F., O.L., M.N., and M.B. designed the experiments. O.L. designed and simulated the light absorption layer. K.G. fabricated the devices. M.S. designed and built the digital mirror device module. Y.S. conceived the optical experiments. B.F. designed the phase masks. R.E., and N.O. designed and performed the 3D localization experiments. R.E. and M.B. wrote the manuscript. All authors reviewed and commented on the manuscript.

Supplementary material. Supplementary material document and supplementary video intended for publication have been provided with the submission.

# Supplementary Material

# Programmable thermocapillary shaping of thin liquid films

Ran Eshel[1], Valeri Frumkin[1,§], Matan Nice[1], Omer Luria[1], Boris Ferdman[2,3], Nadav Opatovski[2], Khaled Gommed[1], Maxim Shusteff[4], Yoav Shechtman[2,3], and Moran Bercovici[1,2,3,*]

[1] Faculty of Mechanical Engineering, Technion - Israel Institute of Technology, Haifa, 3200003 Israel

[2] Department of Biomedical Engineering, Technion - Israel Institute of Technology, Haifa, 3200003 Israel

[3] Russell Berrie Nanotechnology Institute, Technion - Israel Institute of Technology, Haifa, 3200003 Israel

[4] Lawrence Livermore National Laboratory, 7000 East Ave, Livermore, CA 94550, USA

[§] Current affiliation: Department of Mathematics, Massachusetts Institute of Technology, Cambridge, MA, 02139, United States

# Table of Contents

## S.1. Additional experimental data

### Characterization

Figure S1 provides a characterization of the local deformation induced by the projection of an isolated circular spot, as a function of the projection intensity, and for several different geometric configurations of the metallic pads. In all cases, the system was allowed to achieve a steady state, at which point the polymer film was solidified and its surface was measured using a digital holographic microscope (DHM) [1]. Figure S1 presents a cross-section of the resulting axisymmetric film deformation, relative to an initial film thickness of 60 µm, as a function of the illumination intensity. Here, the diameter of the illumination spot is 500 µm, and the absorbing metal pad array consists of 20 x 20 $\mu$m pads with an edge-to-edge distance of 80 µm. As expected, the magnitude of the resulting deformation is proportional to the illumination intensity, and this correlation is linear with a maximum deformation of approximately 1.2 µm, as shown in the inset. The width of the deformation also increases with intensity, yet more moderately, from ~0.6 mm to ~0.8 mm. Further increase in deformation magnitude can be achieved by increasing the pad array density. Figure S1b shows the resulting deformation as a function of the spacing between the absorbing metal array pads, for the same spot size at a fixed (maximum) projection intensity. The results show that a higher pad density leads to absorbance of a larger fraction of light, resulting in increased heating and larger deformation. The inset shows a linear correlation between the area fraction of the surface covered with metal pads and the deformation magnitude, reaching a maximum of approximately 3 $\mu$m. Interestingly, the width of the deformation over this range remains nearly constant. Figure S1c provides information on the repeatability of the resulting deformation by projecting the same spot size and intensity in three independent experiments using the maximum intensity of the light source. Figure S1d presents an isometric view and cross-sections of the resulting deformation for a 500 µm spot size. As expected, for a circular spot, the resulting deformation is axisymmetric.

Figure S2 presents the deformation due to projection of an annulus pattern, with a linear increase in projection intensity along its azimuth, providing a characterization of the achievable deformation resolution. Figure S2a presents the solidified polymer topography as measured by the DHM. Figure S2b presents the projection intensity (blue line) and the resulting deformation magnitude (red circles), along the azimuthal coordinate along path indicated by the red dashed circle in Figure S2a. We find that the deformation correlates well with the light intensity, and the deformation is able to follow a discontinuity in the in the light intensity with a slope of 5.41 µm/mm. Figure S2c presents the cross sections of the deformation at the azimuthal coordinates indicated by corresponding gray lines in Figure S2a. The green squares in Figure S2b show the average over the full width at half maximum (FWHM) of these deformation profiles as a function of the azimuthal coordinate. The direct dependence of the width on the intensity is shown in Figure S3d.

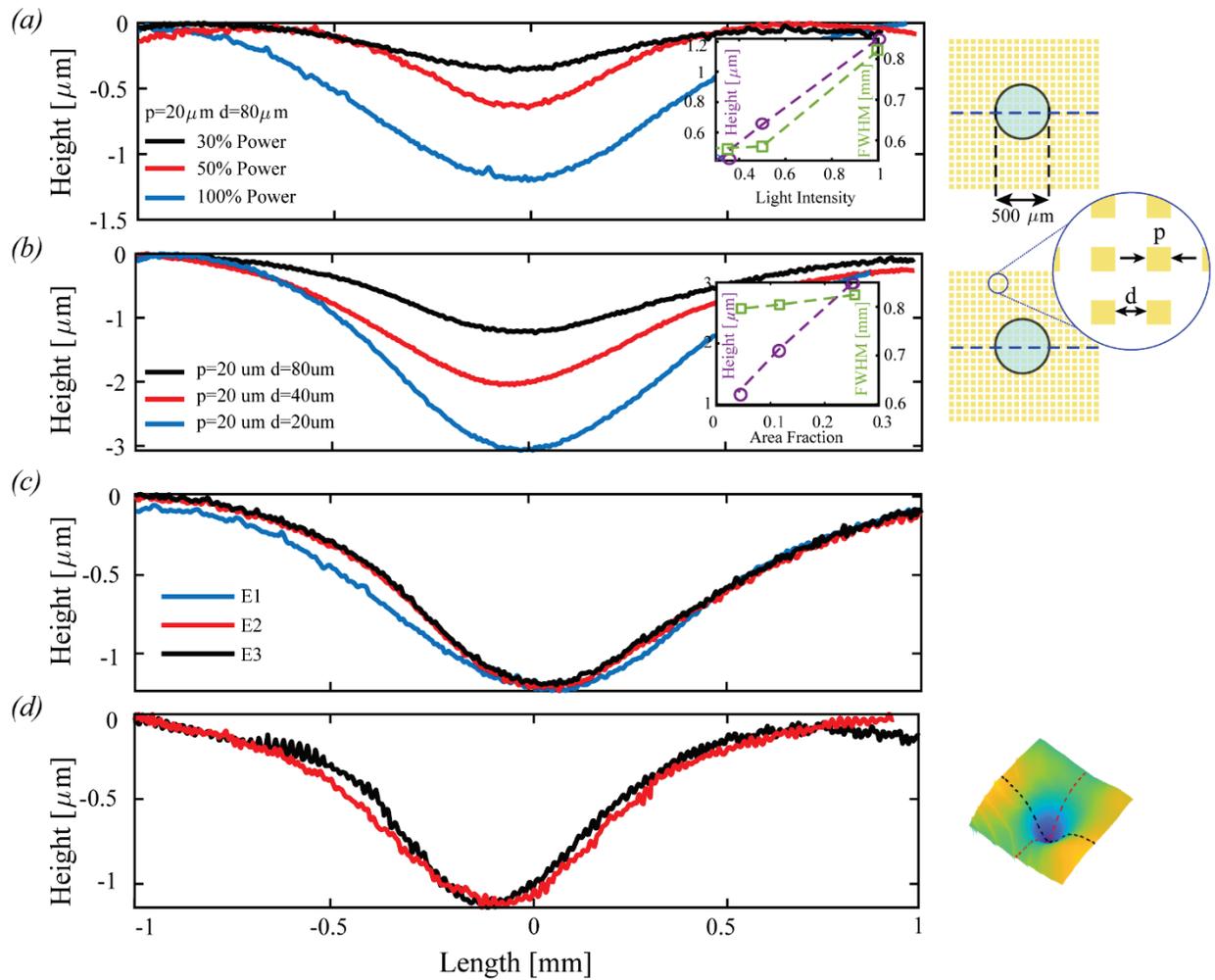

***Figure S1 | Experimental characterization of the deformation magnitude and shape.*** *(a) A cross-section of the deformation along the horizontal axis (as indicated by the dashed blue line) as a function of the illumination intensity, for a fix pad geometry (p = 20 μm, d = 80 μm). As shown in the inset, the deformation magnitude increases linearly with the illumination intensity. The width of the deformation shows a moderate increase between 50% and 100% intensity. (b) A cross section of the deformation along the horizontal axis as a function of the edge-to-edge distance between pads (d=20,40,80 μm), for a fixed illumination intensity (100%) and a fixed pad size (p x p = 20 x 20 μm). As the gap decreases, a larger fraction of the light per unit area is absorbed, allowing further increase in deformation magnitude. Under these conditions, the width of the deformation is essentially unaffected by the gap size. (c) A cross section of the deformation for three repeats at the same intensity. (d) A horizontal and vertical cross section of the deformation, showing good agreement. All the results in the figure were obtained using a 60 μm film thickness and a 500 μm diameter disk-shaped projection.*

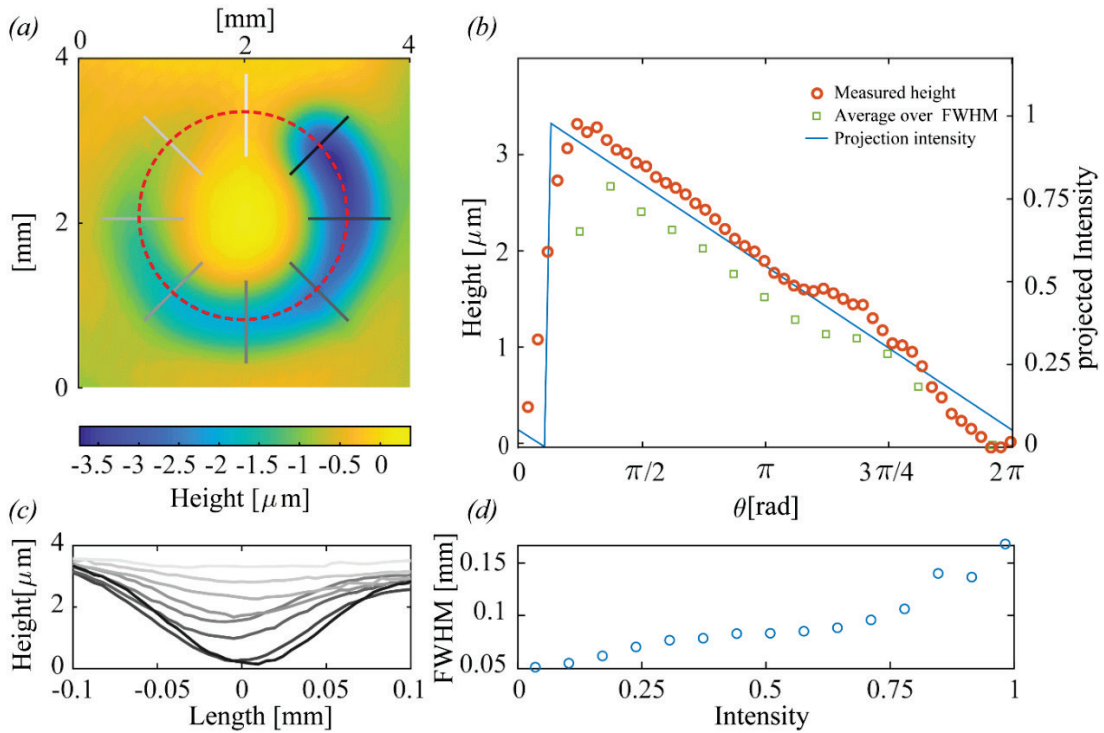

*Figure S2 | Characterization of deformation resolution as a function of projection intensity. (a) Experimental result showing the spatial deformation resulting from projection of an annulus of width 200 μm with an azimuthally varying intensity from 0 to 100%. (b) The magnitude of the deformation (red circles) and the width (FWHM, green squares) at each azimuthal coordinate, together with the projection intensity (blue line) along the centerline of the annulus (red dashed line in a). (c) Radial cross-sections of the deformation at different azimuthal locations as indicated by corresponding grayscale lines in a. (d) Full width at half maximum of the deformation as a function of intensity along the annulus pattern.*

## Surface roughness measurements

Figure S3 presents a representative atomic force microscopy (AFM) measurement that we performed over an area of 2.2 x 2.2 μm in order to evaluate the surface roughness of the solidified polymer surface. The resulting topography shows a surface roughness of less than 1 nm RMS over the measured area, which is excellent for nearly all optical applications.

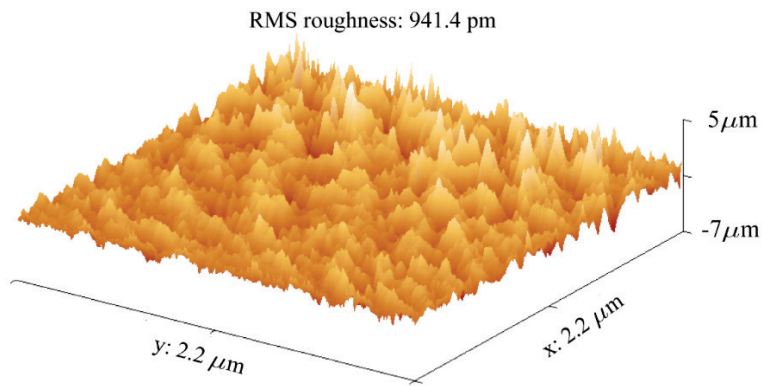

*Figure S3 | AFM measurement of the surface of the solidified polymer. The RMS of the surface roughness is less than 1 nm.*

## S.2. Solutions to the inverse problem

We are interested in solving the inverse problem, namely, finding the temperature field required to create a desired deformation.

The differential equation describing the steady state deformation of a thin liquid film subjected to a non-uniform temperature gradient is given by (see Appendix A for detailed derivation)

$$\vec{\nabla} \cdot \left[ \frac{S}{3} H^3 \vec{\nabla} \nabla^2 H - \frac{G}{3} H^3 \vec{\nabla} H - \frac{1}{2} H^2 \vec{\nabla} \left( \frac{\vartheta_b + \Theta}{1 + BH} \right) \right] = 0, \tag{S1}$$

where $H$ is the dimensionless distance of the free surface from the flat rigid floor, $\vartheta_b$ is the temperature distribution on the bottom surface, $S$ is the inverse capillary number, $G$ is the gravitation number, $B$ is the Biot number, and $\Theta$ is the scaled magnitude of the temperature at infinity (see details in the Appendix).

### One dimensional analysis

We start by obtaining a solution to the inverse problem for the one-dimensional case, described by

$$\partial_X \left[ \frac{S}{3} H^3 \partial_X^3 H - \frac{G}{3} H^3 \partial_X H - \frac{1}{2} H^2 \partial_X \left( \frac{\vartheta_b + \Theta}{1 + BH} \right) \right] = 0, \tag{S2}$$

where $X$ is the non-dimensional spatial coordinate. We assume that the temperature variations occur far from the boundaries, $X = \pm \mathcal{L}$, and therefore at the boundaries the film can be assumed to be flat.

$$H(\pm \mathcal{L}) = 1, \quad \partial_X^n H(\pm \mathcal{L}) = 0 \; \{n = 1,2,3\}, \quad \vartheta_b(\pm \mathcal{L}) = -\Theta, \quad \partial_X \vartheta_b(\pm \mathcal{L}) = 0. \tag{S3}$$

Integrating equation (S2) in $X$ and dividing by $H^2$ we obtain

$$\frac{S}{3} H \partial_X^3 H - \frac{G}{3} H \partial_X H - \frac{1}{2} \partial_X \left( \frac{\vartheta_b + \Theta}{1 + BH} \right) = \frac{C_0}{H^2}, \tag{S4}$$

where $C_0$ is an integration constant. The *LHS* can be integrated by parts, and vanishes when using the boundary conditions (S3),

$$\int_{-\mathcal{L}}^{\mathcal{L}} \left[ \frac{S}{3} H \partial_X^3 H - \frac{G}{3} H \partial_X H \frac{G}{3} H \partial_X H - \frac{1}{2} \partial_X \left( \frac{\vartheta_b + \Theta}{1 + BH} \right) \right] dX = 0 \tag{S5}$$

Since the integration of the *RHS* of (S4) is non-zero, the coefficient $C_0$ must vanish, $C_0 = 0$.

The 1D steady state equation (S4) can thus be simplified to

$$\frac{S}{3} H \partial_X^3 H - \frac{G}{3} H \partial_X H - \frac{1}{2} \partial_X \left( \frac{\vartheta_b + \Theta}{1 + BH} \right) = 0. \tag{S6}$$

Integrating equation (S6) to an arbitrary position $\xi$ yields an explicit expression for the temperature variation, $\theta_b(\xi)$, required in order to create the desired deformation $H(\xi)$.

$$\vartheta_b(\xi) = \frac{1 + BH(\xi)}{3} \left[ 2SH(\xi) \partial_\xi^2 H(\xi) - S \left( \partial_\xi H(\xi) \right)^2 - G(H(\xi)^2 - 1) \right] - \Theta. \tag{S7}$$

### Two-dimensional analysis

The two-dimensional case cannot be directly integrated. We therefore propose an ansatz based on the 1D solution,

$$\vartheta_b = \frac{1+BH}{3}\left[2SH\nabla^2 H - S(\vec{\nabla}H)^2 - G(H^2-1)\right] - \Theta, \tag{S8}$$

or in dimensional form,

$$\theta_b - \theta_\infty = \frac{1+\frac{\hbar}{k}h}{3\beta}\left[2\sigma_r h\nabla^2 h - \sigma_r(\vec{\nabla}h)^2 - \rho g(h^2 - h_0^2)\right]. \tag{S9}$$

To test the validity of the proposed ansatz, we reconstruct a differential equation that yields this solution (S8), and check to what extent it deviates from (S1). Equation (S8) can be recast as

$$\frac{1}{2}\left(\frac{\vartheta_b + \Theta}{1+BH}\right) = \frac{1}{6}\left(2SH\nabla^2 H - S(\vec{\nabla}H)^2 - G(H^2 - 1)\right), \tag{S10}$$

and taking its gradient then yields

$$\frac{1}{2}\vec{\nabla}\left(\frac{\vartheta_b + \Theta}{1+BH}\right) = \frac{1}{3}\left(SH\vec{\nabla}\nabla^2 H + S\vec{\nabla}H\nabla^2 H - \frac{1}{2}S\vec{\nabla}(\vec{\nabla}H)^2 - GH\vec{\nabla}H\right). \tag{S11}$$

To compare equation (S11) to (S1), we note that equation (S1) can be written as $\vec{\nabla}\cdot(H^2\vec{F}) = 0$, which can be satisfied by

$$\vec{F} = \frac{S}{3}H\vec{\nabla}\nabla^2 H - \frac{G}{3}H\vec{\nabla}H - \frac{1}{2}\vec{\nabla}\left(\frac{\vartheta_b + \Theta}{1+BH}\right) = 0, \tag{S12}$$

or

$$\frac{1}{2}\vec{\nabla}\left(\frac{\vartheta_b + \Theta}{1+BH}\right) = \frac{S}{3}H\vec{\nabla}\nabla^2 H - \frac{G}{3}H\vec{\nabla}H. \tag{S13}$$

Subtracting (S11) from (S13), we are left only with the second and third terms on the RHS of (S11),

$$0 = \frac{S}{6}\left[\vec{\nabla}(\vec{\nabla}H)^2 - 2\vec{\nabla}H\nabla^2 H\right]. \tag{S14}$$

The RHS in (S14) can thus be regarded as the difference between the approximate (ansatz) solution and the exact solution (S1),

$$\vec{\epsilon} = \frac{S}{6}\left[\vec{\nabla}[(\partial_X H)^2 + (\partial_Y H)^2] - 2\vec{\nabla}H(\partial_X^2 H + \partial_Y^2 H)\right], \tag{S15}$$

which can be written explicitly in terms of its components as

$$\vec{\epsilon} = \frac{S}{3}[(\partial_Y H \partial_{XY}^2 H - \partial_X H \partial_Y^2 H)\,,\,(\partial_X H \partial_{XY}^2 H - \partial_Y H \partial_X^2 H)] \tag{S16}$$

The ansatz solution is thus exact when the non-normalized Gaussian curvature of the surface, $K = \partial_X^2 H \partial_Y^2 H - (\partial_{XY}^2 H)^2$, vanishes. For other surfaces, the error would depend on the magnitude of the Gaussian curvature, and we provide a numerical assessment for this error in the following section.

## S.3. Numerical verification of the two-dimensional inverse problem solution

We seek a solution that will satisfy equation (S12),

$$F_x = \frac{S}{3}H\partial_X \nabla^2 H - \frac{G}{3}H\partial_X H - \frac{1}{2}\partial_X \left(\frac{\vartheta_b + \Theta}{1 + BH}\right) = 0, \quad (S17.a)$$

$$F_y = \frac{S}{3}H\partial_Y \nabla^2 H - \frac{G}{3}H\partial_Y H - \frac{1}{2}\partial_Y \left(\frac{\vartheta_b + \Theta}{1 + BH}\right) = 0. \quad (S17.b)$$

By defining $\mathrm{T} = \frac{\vartheta_b + \Theta}{1 + BH}$, the equations can be expressed as

$$\partial_X \mathrm{T} = \frac{2}{3}H[S\partial_X(\nabla^2 H) - G\partial_X H], \quad (S18.a)$$

$$\partial_Y \mathrm{T} = \frac{2}{3}H[S\partial_Y(\nabla^2 H) - G\partial_Y H]. \quad (S18.b)$$

We specify a desired deformation map $H(X,Y)$ and integrate equation (S18.a) numerically in $X$ from 0 to an arbitrary coordinate within the domain, $\xi$. The solution can be thus expressed as

$$\mathrm{T} = A_1(\xi, Y) + A_2(Y), \quad (S19)$$

where $A_1 = \int_{-\mathcal{L}}^{\xi} \frac{2}{3}H[S\partial_X(\nabla_2^2 H) - G\partial_X H]dX$ is the result of the numerical integration and $A_2(Y)$ is an unknown function. Substituting (S19) into equation (S18.b) yields an equation for A2,

$$\frac{d}{dY}A_2(Y) = \frac{2}{3}H[S\partial_Y(\nabla^2 H) - G\partial_Y H] - \partial_Y A_1(X, Y), \quad (S20)$$

which we integrate numerically in $Y$ between $-\mathcal{L}$ and an arbitrary coordinate within the domain, $\eta$, to find $A_2$. The temperature field is then given by

$$\vartheta_b = (1 + BH)(A_1 + A_2) - \Theta \quad (S21)$$

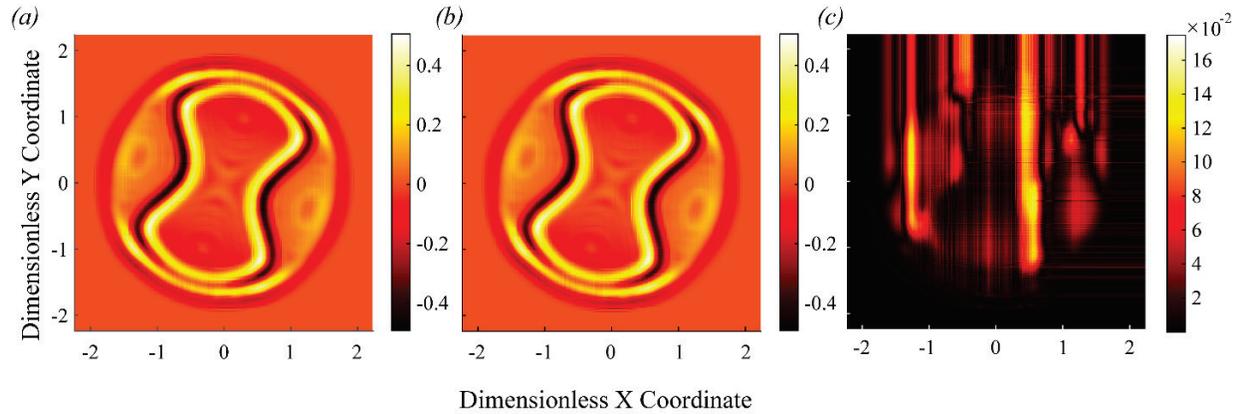

***Figure S4 | A comparison between analytical and numerical solutions to the inverse problem of a tetrapod phase mask.** (a) The dimensionless temperature based on the analytical solution S8. (b) The dimensionless temperature based the numerical solution S21. (c) The relative difference (in percent) between the analytical and numerical solutions, showing that the deviation is less than 0.17 %.*

Figure S4 shows a comparison between the resulting dimensionless temperature obtained from the analytical solution (Figure S4a, corresponding to eq. S8) and from the numerical solution (Figure S4b, corresponding to eq. S21) for the case of the 3D localization microscopy mask ('tetrapod mask') presented in Figure 5 of the main manuscript. For clarity of presentation, we subtracted from both solutions the mean temperature of the boundaries, so that the dimensionless temperature outside the mask area is set to zero. We defined the relative difference between the solutions as the absolute difference between them divided by the amplitude of temperature, and present the results in Figure S4c. As can be seen, the maximal relative difference is approximately 0.17 %.

## S.4. Material properties

Table S1. Material properties that were used in all of the analytical and numerical solutions.

| Property | Value | Units | Source |
|---|---|---|---|
| $L_x, L_y$ | $6 \cdot 10^{-3}$ | $m$ | Chamber size |
| $h_0$ | $60 \cdot 10^{-6}$ | $m$ | Calculated film thickness based on chamber size and injected volume |
| $\sigma_r$ | 28.5 | $\frac{mN}{m}$ | Based on the properties of the polymer CPS1050 |
| $\beta$ | 0.866 | $\frac{mN}{mK}$ | Based on the properties of the polymer CPS1050 |
| $\rho$ | 1200 | $\frac{kg}{m^3}$ | Based on the properties of NOA63 |
| $\mu$ | 2 | $Pa \cdot s$ | Based on the properties of NOA63 |
| $C_p$ | 1632.852 | $\frac{J}{kgK}$ | Based on the properties of PSF-5cSt oil |
| $k$ | $117.152 \cdot 10^{-3}$ | $\frac{W}{mK}$ | Based on the properties of PSF-5cSt oil |
| $g$ | 9.81 | $\frac{m}{s^2}$ | Earth's gravity acceleration |
| $\hbar$ | 10 | $\frac{W}{m^2 K}$ | Estimation based on common natural convection coefficients |
| $\theta_{b_{min}}, \theta_{b_{max}}$ | 23, 25 | °C | Assumed. Used for scaling only. |
| $\overline{\theta_b}$ | 24 | °C | Assumed. Used for scaling only. |
| $\theta_\infty$ | 23 | °C | Assumed. Used for scaling only. |
| $l$ | $2 \cdot 10^{-3}$ | $m$ | Assumed. Used for scaling only. |
| $\nu = \frac{\mu}{\rho}$ | $1.7 \cdot 10^{-3}$ | $\frac{m^2}{s}$ | Calculated |
| $\alpha = \frac{k}{\rho C_p}$ | $5.979 \cdot 10^{-8}$ | $\frac{m^2}{s}$ | Calculated |
| $\Delta\theta = \theta_{b_{max}} - \theta_{b_{min}}$ | 2 | °C | Calculated |

# Appendix A: Derivation of the steady-state equation for the shape of a thin liquid film subjected to a non-uniform temperature distribution

We follow the derivation first presented by Vanhook et. al [2] and derive a 2D steady state equation for the free surface of a liquid film under the action of the thermocapillary effect. As illustrated in Figure S5, we consider a liquid with density $\rho$, dynamic viscosity $\mu$, surface tension $\sigma$, thermal conductivity $k$, and heat capacity $C_p$, filling a shallow square-shaped cavity of length $L$, to an initial height $h_0$.

A small subarea of the substrate with a characteristic length $l$ is heated with a prescribed temperature $\theta_b(x,y)$, which is conducted to the free surface, yielding a non-uniform surface temperature. This yields surface tension gradients, which in turn lead to the transfer of mass along the interface, known as the thermocapillary effect [3]. Throughout the analysis we assume that the film thickness $h_0$ is significantly smaller than the characteristic length $l$, so that $\varepsilon = \frac{h_0}{l} \ll 1$.

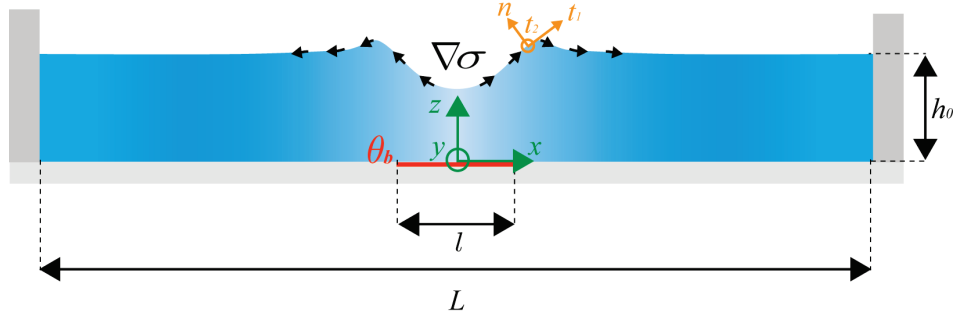

**Figure S5 | Schematic description of the problem under consideration.** *A thin liquid film with an initial height $h_0$, rests in a shallow, square-shaped cavity with a side length L. A small subarea of the substrate with a characteristic length l is heated with a prescribed temperature $\theta_b(x,y)$. The resulting non-uniform temperature distribution at the free surface induces gradients in surface tension, which, through the thermocapillary effect, lead to deformations of the liquid film.*

## Governing equations and boundary conditions in dimensional form

We separate our problem into a hydrodynamic problem and a thermal one, and begin with the prior. The governing equations are the continuity equation and steady-state Navier-Stokes (NS) equations,

$$\rho(\boldsymbol{\nabla} \cdot \boldsymbol{U}) = 0, \qquad (A1.a)$$

$$\rho \boldsymbol{U} \cdot \vec{\nabla} \boldsymbol{U} = -\vec{\nabla} p + \mu \nabla^2 \boldsymbol{U} + \rho \boldsymbol{g}, \qquad (A1.b)$$

where $\boldsymbol{U}$ is the three-dimensional velocity field, $\boldsymbol{g} = (0,0,-g)$ is the gravity acceleration vector, and $p$ is the pressure.

At $z = 0$ the liquid is subject to no-slip and no-penetration conditions,

$$\boldsymbol{U}(z=0) = 0. \qquad (A2)$$

At the liquid-air interface, $z = h(x,y)$, the normal component of the velocity satisfies the kinematic boundary condition

$$w = \boldsymbol{u_2} \cdot \nabla h, \qquad (A3.a)$$

where $h = h(x,y)$ is the position of the free surface, $\boldsymbol{u_2} = (u,v)$ is the planar velocity vector, $\nabla = (\partial_x, \partial_y)$ is the 2D gradient in the $x$ and $y$ directions, and $w$ is the velocity component in the $z$ direction.

Another boundary condition is a stress balance at the free surface,

$$(\overline{T} - \overline{T}_{air}) \cdot \mathbf{n} = 2\mathcal{H}\sigma\mathbf{n} + \partial_{t_1}\sigma \cdot \mathbf{t_1} + \partial_{t_2}\sigma \cdot \mathbf{t_2}. \quad (A3.b)$$

Here $\partial_{t_i}$ is the directional derivative in the $t_i$ direction ($i = 1,2$), $\overline{T}$ and $\overline{T}_{air}$ are the stress tensors in the liquid and air, respectively, $\mathbf{n}$ and $\mathbf{t_i}$ are the normal and tangential unit vectors, and $2\mathcal{H}$ is the mean curvature.

### Scaling and asymptotics

We assume that the velocities in the $x, y$ directions are of similar magnitude, and introduce the following scaling

$$U = \frac{u}{U_0}, V = \frac{v}{U_0}, W = \frac{w}{W_0}, X = \frac{x}{l}, Y = \frac{y}{l}, Z = \frac{z}{h_0},$$
$$H = \frac{h}{h_0}, P = \frac{p}{P_0}, G = \frac{\rho h_0}{P_0}g, \Sigma = \frac{\sigma}{\Delta\sigma}, \quad (A4)$$

where $\Delta\sigma$ is a characteristic change in the surface tension resulting from temperature gradients, $\Delta\sigma = \beta\Delta\theta$. The dimensionless governing equations are hence:

$$\varepsilon U_0 \partial_X U + \varepsilon U_0 \partial_Y V + W_0 \partial_Z W = 0, \quad (A5.a)$$

$$\varepsilon \text{Re}\left(U\partial_X U + V\partial_Y U + \frac{W_0}{U_0\varepsilon} W\partial_Z U\right) = -\varepsilon \frac{h_0 P_0}{U_0 \mu} \partial_X P + (\varepsilon^2 \partial_X^2 U + \varepsilon^2 \partial_Y^2 U + \partial_Z^2 U),$$

$$\varepsilon \text{Re}\left(U\partial_X V + V\partial_Y V + \frac{W_0}{U_0\varepsilon} W\partial_Z V\right) = -\varepsilon \frac{h_0 P_0}{U_0 \mu} \partial_Y P + (\varepsilon^2 \partial_X^2 V + \varepsilon^2 \partial_Y^2 V + \partial_Z^2 V), \quad (A5.b)$$

$$\varepsilon \text{Re} \frac{W_0}{U_0}\left(U\partial_X W + V\partial_Y W + \frac{W_0}{U_0\varepsilon} W\partial_Z W\right) = -\frac{h_0 P_0}{U_0 \mu} \partial_Z P + \frac{W_0}{U_0}(\varepsilon^2 \partial_X^2 W + \varepsilon^2 \partial_Y^2 W + \partial_Z^2 W) - \frac{h_0 P_0}{U_0 \mu} G,$$

where $\text{Re} = \frac{\rho U_0 h_0}{\mu}$ is the Reynolds number and $U_0$ will be defined in the following section. From the continuity equation (A5.a) we obtain the proper scaling for the velocity in the $z$ direction, $W_0 = \varepsilon U_0$ and from the momentum equations (A5.b) we obtain the proper scaling for the pressure, $P_0 = \frac{\mu U_0}{\varepsilon h_0}$.

Projecting boundary condition (A3.b) onto the first tangential vector $\mathbf{t_1}$ yields the equation

$$\varepsilon P_0 (\partial_Y H \partial_Z U - \partial_X H \partial_Z V) = \frac{\Delta\sigma}{l}(\partial_X \Sigma \partial_Y H - \partial_Y \Sigma \partial_X H). \quad (A6)$$

A similar result is achieved by projecting (A3.b) onto the second tangential vector, $\mathbf{t_2}$. We substitute the definition of $P_0$ and rewrite (A6) as

$$\partial_Y H\left(\partial_Z U - \frac{\varepsilon\Delta\sigma}{\mu U_0}\partial_X \Sigma\right) - \partial_X H\left(\partial_Z V - \frac{\varepsilon\Delta\sigma}{\mu U_0}\partial_Y \Sigma\right) = 0. \quad (A7)$$

Since we require the balance in (A7) to hold independently of the derivatives of $H$, we obtain the proper scaling for the velocity, $U_0 = \varepsilon \Delta\sigma/\mu$ and the dimensionless boundary conditions (A11.c).

In the normal direction, equation (A3.b) balances the external pressure with the surface tension. We assume that the overall change in surface tension, $\Delta\sigma$, is small relative to a reference surface tension $\sigma_r$ and will further discuss this in the thermal section. Considering the aforementioned assumption, we can write the surface tension as $\sigma \approx \sigma_r$ and by projecting boundary condition (A3.b) onto the normal vector $\boldsymbol{n}$, the normal stress balance condition in dimensionless form can be expressed as

$$-P_0 P = \frac{\varepsilon \sigma_r}{l}(\partial_X^2 H + \partial_Y^2 H). \tag{A8}$$

Substituting the expression for $P_0$ we can rewrite equation (A8) as

$$-P = S(\partial_X^2 H + \partial_Y^2 H), \tag{A9}$$

where $S = \varepsilon^3 \frac{\sigma_r}{\mu U_0} = \varepsilon^2 \frac{\sigma_r}{\Delta \sigma}$ is the surface tension number. For our case $\frac{\sigma_r}{\Delta \sigma} = 10^2 \sim 10^3$ and $\varepsilon^2 = O(10^{-3})$ therefore $S = O(1)$ and the balance is maintained.

With the obtained scaling, we now turn to asymptotic analysis of the NS equations (A5.b). Noting that for the typical values in our system $Re \ll 1$, the leading order of the equations take the form

$$\partial_Z^2 U = \partial_X P, \qquad \partial_Z^2 V = \partial_Y P, \tag{A10.a}$$

$$\partial_Z P = -G, \tag{A10.b}$$

with the boundary conditions at $Z = 0$:

$$U = V = W = 0 \tag{A11}$$

and at $Z = H$:

$$W = U \partial_X H + V \partial_Y H, \tag{A12.a}$$

$$\partial_Z U = \partial_X \Sigma, \qquad \partial_Z V = \partial_Y \Sigma, \tag{A12.b}$$

$$-P = S[\partial_X^2 H + \partial_Y^2 H]. \tag{A12.c}$$

Integrating the continuity equation (A5.a) in the Z direction, applying Leibniz's rule and using the boundary conditions (A11) and (A12.a) we obtain an integral form of the kinematic condition

$$\partial_X \int_0^H U dZ + \partial_Y \int_0^H V dZ = 0. \tag{A13}$$

Solving for the velocity field as a function of the pressure gradient (equations (A10.a) with B.C.s (A11) and (A12.b)), plugging it into equation (A13), and performing the integration yields

$$\left[\partial_X \left(-\frac{\partial_X P H^3}{3} + \frac{\partial_X \Sigma H^2}{2}\right) + \partial_Y \left(-\frac{\partial_Y P H^3}{3} + \frac{\partial_Y \Sigma H^2}{2}\right)\right]_{Z=H} = 0. \tag{A14}$$

Equation (A14) can be compactly written as

$$\vec{\nabla} \cdot \left[\frac{H^2}{2}\vec{\nabla}\Sigma - \frac{H^3}{3}\vec{\nabla}P\right] = 0, \tag{A15}$$

where $\vec{\nabla} = (\partial_X, \partial_Y)$.

Integrating equation (A10.b) with B.C. (A12.c) allows us to express the pressure gradient in terms of the dimensionless free surface,

$$\vec{\nabla}P = G\vec{\nabla}H - S\vec{\nabla} \cdot \nabla^2 H. \qquad (A16)$$

Substituting (A16) into (A15) results in the final form of the equation for the steady-state position of the free surface,

$$\vec{\nabla} \cdot \left[ \frac{H^2}{2} \vec{\nabla}\Sigma - \frac{H^3}{3}(G\vec{\nabla}H - S\vec{\nabla} \cdot \nabla^2 H) \right] = 0. \qquad (A17)$$

## Thermocapillary effect

In order to evaluate $\vec{\nabla}\Sigma$ we proceed to solve the coupled thermal problem, which is governed by the steady state energy equation,

$$\boldsymbol{U} \cdot \vec{\nabla}\theta = \alpha(\partial_x^2 \theta + \partial_y^2 \theta + \partial_z^2 \theta), \qquad (A18)$$

where $\theta$ is the liquid's temperature and $\alpha = \frac{k}{\rho C_p}$ is its thermal diffusivity.

At $z = 0$ the liquid is subject to a prescribed temperature profile $\theta_b(x,y)$:

$$\theta(z = 0) = \theta_b(x,y), \qquad (A19)$$

and the heat transfer at the liquid-air interface, $z = h(x,y)$, is described by Newton's cooling law:

$$k\vec{\nabla}\theta_h \cdot \boldsymbol{n} + \hbar(\theta_h - \theta_\infty) = 0, \qquad (A20)$$

where $\hbar$ is the heat transfer coefficient and $\theta_\infty$ is the ambient gas temperature.

We scale the temperature by $\vartheta = \frac{\theta - \bar{\theta}_b}{\Delta\theta}$, where $\Delta\theta = \theta_{b_{max}} - \theta_{b_{min}}$, and $\bar{\theta}_b$, $\theta_{b_{max}}$, $\theta_{b_{min}}$ are the average temperature at the base, and its maximum and minimum values, respectively.

Equation (A18) in dimensionless form is thus

$$\varepsilon \text{Pe}(U\partial_X \vartheta + V\partial_Y \vartheta + W\partial_Z \vartheta) = \varepsilon^2 \partial_X^2 \vartheta + \varepsilon^2 \partial_Y^2 \vartheta + \partial_Z^2 \vartheta, \qquad (A21)$$

where $\text{Pe} = \frac{U_0 h_0}{\alpha}$ is the thermal Peclet number. We note that in our case $\text{Pe} = \frac{\varepsilon \Delta \sigma h_0}{\alpha \mu} = O(10^{-3}) \ll 1$.

We proceed to write $\vartheta$ as an asymptotic expansion in $\varepsilon$, which at the leading order yields

$$\partial_Z^2 \vartheta = 0, \qquad (A22)$$

with the boundary conditions

$$\begin{aligned} Z = 0: &\quad \vartheta_b = \frac{\theta_b - \bar{\theta}_b}{\Delta\theta}, \\ Z = H(X,Y): &\quad [\partial_Z \vartheta + B(\vartheta + \Theta)]_{Z=H} = 0, \end{aligned} \qquad (A23)$$

where $\vartheta_b$ is the dimensionless temperature distributions at the bottom surface, $\Theta = \frac{\bar{\theta}_b - \theta_\infty}{\Delta\theta}$ is a ratio between temperature differences, and $B = \frac{\hbar h_0}{k}$ is the Biot number.

Equation (A22) with B.C.s (A23) yields the temperature distribution inside the liquid film

$$\vartheta = \vartheta_b - \frac{(\vartheta_b + \Theta)B}{1 + BH} Z, \qquad (A24)$$

and hence $\vartheta|_{Z=H} = \frac{\vartheta_b - BH\Theta}{1+BH}$.

We assume that the surface depends solely on the temperature at the interface $\sigma = \sigma(\theta(x, y, z = h))$, and that for small variations in temperature the surface tension changes linearly such that

$$\Delta\sigma = -\beta\Delta\theta_{z=h}, \qquad (A25)$$

where $\beta$ is a positive constant and a property of the material and $\Delta\theta_{z=h}$ is the variation in temperature at the interface. The negative sign on the RHS of (A25) represents a decrease in surface tension as the temperature increases. This is consistent with our previous assumption of small variations in surface tension since for the liquid we used $\frac{\beta}{\sigma_r} \approx 10^{-3} \frac{1}{°K}$ and for our system $\Delta\theta_h \approx 1°K$.

Using the chain rule, its gradient can be written as:

$$\vec{\nabla}\sigma = \frac{d\sigma}{d\theta}(\vec{\nabla}\theta_{z=h} + \Delta\theta\partial_z\theta_{z=h}\vec{\nabla}h). \qquad (A26)$$

In dimensionless form the surface tension gradient is

$$\vec{\nabla}\Sigma = -(\vec{\nabla}\vartheta_{Z=H} + \partial_Z\vartheta_{Z=H}\vec{\nabla}H), \qquad (A27)$$

Substituting equation (A24) into (A27) we obtain

$$\vec{\nabla}\Sigma = -\vec{\nabla}\left(\frac{\vartheta_b + \Theta}{1 + BH}\right). \qquad (A28)$$

Further substituting equation (A28) into equation (A17) yields the final equation defining the relation between the prescribed temperature and the resulting surface topography:

$$\nabla \cdot \left[\frac{S}{3}H^3\vec{\nabla}\nabla^2 H - \frac{G}{3}H^3\vec{\nabla}H - \frac{1}{2}H^2\vec{\nabla}\left(\frac{\vartheta_b + \Theta}{1 + BH}\right)\right] = 0. \qquad (A29)$$